\documentclass[aps,prl,reprint,showpacs,twocolumn,longbibliography,superscriptaddress]{revtex4-1}

\usepackage{graphicx}
\usepackage{dcolumn}
\usepackage{bm}
\usepackage{braket}
\usepackage{leftidx}
\usepackage{cancel}
\usepackage{amsmath}
\usepackage{epstopdf}
\usepackage{color}
\usepackage[mathcal]{eucal}
\usepackage{float}
\usepackage{url}
\usepackage{hyperref}

\usepackage{dcolumn}
\usepackage{bm}

\begin{document}


\title{Ultra-large actively tunable photonic band gaps \\ via plasmon-analog of index enhancement}

\author{Emre Y\"{u}ce}
\affiliation{Department of Physics, Middle East Technical University, 06800 Ankara, Turkey}
\affiliation{Center for Solar Energy Research and Applications (G\"{U}NAM), Middle East Technical University, Dumlupinar Blvd. 1, 06800 Ankara, Turkey}

\author{Ahmet Kemal Demir}
\affiliation{Department of Physics, Bilkent University, 06800 Ankara, Turkey}

\author{Zafer Artvin}
\affiliation{Institute  of  Nuclear  Sciences, Hacettepe University, 06800 Ankara, Turkey}
\affiliation{Department of Nanotechnology and Nanomedicine, Hacettepe University, Ankara, Turkey}
\affiliation{Central Laboratory, Middle East Technical University, 06800 Ankara, Turkey}

\author{Ramazan Sahin}
\affiliation{Faculty of Science, Department of Physics, Akdeniz University, 07058 Antalya, Turkey}

\author{Alpan Bek}
\affiliation{Department of Physics, Middle East Technical University, 06800 Ankara, Turkey}
\affiliation{Center for Solar Energy Research and Applications (G\"{U}NAM), Middle East Technical University, Dumlupinar Blvd. 1, 06800 Ankara, Turkey}

\author{Mehmet Emre Tasgin} 
\affiliation{Institute  of  Nuclear  Sciences, Hacettepe University, 06800 Ankara, Turkey}

\date{\today}

\begin{abstract}
We present a novel method for active continuous-tuning of a band gap which has a great potential to revolutionalize current photonic technologies. We study a periodic structure of x and y-aligned nanorod dimers. Refractive index of a y-polarized probe pulse can be continuously-tuned by the intensity of an x-polarized auxiliary (pump) pulse. Order of magnitude index-tuning can be achieved with a {\it vanishing loss} using the plasmon-analog of refractive index enhancement~[PRB 100, 075427 (2019)]. Thus, a large band gap can be created from a non-existing gap via the auxiliary pulse. We also present a ``proof of principle'' demonstration of the phenomenon using numerical solutions of Maxwell equations. The new method, working for any crystal dimensions, can also be utilized as a {\it linear} photonic switch operating at tens of femtoseconds. 
\end{abstract}

\maketitle



Ability of crafting materials at dimensions comparable to wavelength of light not only allowed control of electromagnetic radiation at nanoscale, but also made development of metamaterials~\cite{liu2011metamaterials}, such as photonic crystals~(PCs)~\cite{yablonovitch1987inhibited,yablonovitch1991photonic, Wijnhoven1998}, possible. PCs, periodically altering refractive index materials, reshape the propagation (dispersion) of the incident light and even can forbid it at certain propagation directions within some frequency ranges, named as photonic band gaps~(PBGs)~\cite{Leistikow2011}. PBGs, whose structure depends on the periodicity and index contrast, can be utilized as waveguides~\cite{knight1996all}, optical cavities~\cite{akahane2003high}, optical memories~\cite{kuramochi.2014} and switches~~\cite{euser2007ultrafast, yuce2013all}. This way PBGs enable on-chip photonic signal processing~\cite{wang2018chip}. These structures enable control even over quantum phenomena~\cite{haroche1989cavity,lodahl2004controlling,madsen2011observation} such as spontaneous emission~\cite{scully2015single,scully2009super} and entanglement~\cite{tasgin2017many} at nanoscale.

A conventional PC ---whose PBG structure remains fixed after manufacturing--- cannot operate, e.g., as a switch, although it is useful for photonic waveguide and cavity applications. External control of the constituent material's refractive index, however, demonstrated to tune the PBG width about a few percent which allows the utilization of PCs as photonic switches. Refractive index of semiconductors (materials commonly employed in current PC technologies owing to their relatively-high index, $n$=3.5-4, and small absorption) can be tuned externally (actively) via optical heating~\cite{reed2004silicon}, free-carrier excitation~\cite{almeida2004all} and electronic Kerr-effect~\cite{ctistis2011ultimate,yuce2013all}. While technique of free-carrier excitation provides the largest index change~($\sim$3\%), its response time~(tens of picoseconds) is much longer compared to the modulation time of the Kerr effect (a few femtosecond). Using the Kerr effect, in contrast, achievable index change is only $\sim$1\%.

Metal nanostructures, which can localize the incident light into nm-sized hotspots, are also utilized for photonic technologies. Hybrid metastructures, created with semiconductors and metallic nanostructures, increase the versatility of integrated photonic structures~\cite{fan2015dynamic,he2019tunable,kang2019recent}. Yet, regarding the modulation depth, response time and (especially) metal-induced losses their functionalities are limited~\cite{fan2015dynamic,he2019tunable,kang2019recent}. 

Therefore, a mechanism providing; (i) a large refractive index change (ii) minimal loss, and (iii) a short response time will revolutionalize the dynamic control of light on-a-chip. Fortunately, in this paper, we propose the utilization of a recently-discovered extraordinary index enhancement (linear control) scheme for achieving a {\it game-changing} control over the PGBs of a PC. Actually, utilization of index enhancement~\cite{ScullyZubairyBook,fleischhauer2005electromagnetically} as PCs has already been studied in cold atoms~\cite{tasgin2007photonic,mustecapliouglu2005photonic}.

\begin{figure}
\centering
\includegraphics[width=0.47\textwidth]{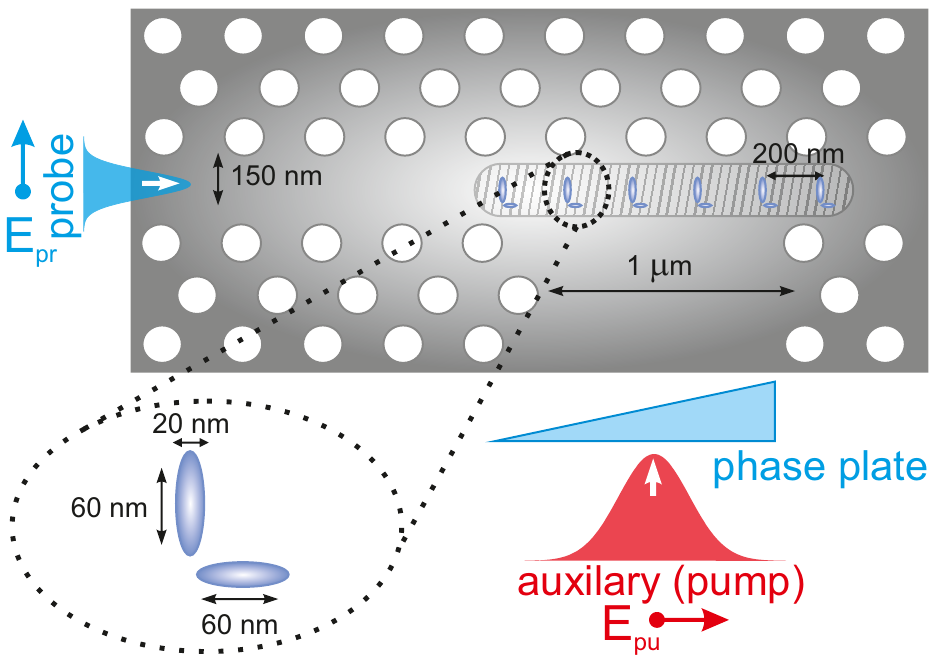} 
\caption{A sample setup for the continuous band gap tuning~(Fig.~\ref{fig4}) and femtosecond photonic switch~(Fig.~\ref{fig5}). Probe~($E_1\hat{\mathbf{ y}}$) and auxiliary~($E_2\hat{\mathbf{x}}$) pulses reach the periodic layers of nanorod dimers via defect waveguides. Intensity and phase of auxiliary pulse controls the presence and width of the photonic bad gap~(Fig.~\ref{fig4}). The phase plate matches the phase-difference between the aux and probe pulses at all dimers.  When an ultrashort auxiliary pulse is employed the same periodic structure behaves as a linear photonic switch~(Fig.~\ref{fig5}).  }
\label{fig1}
\end{figure}

We utilize a recently explored phenomenon: plasmon-analog of index enhancement~\cite{LavrinenkoPRB2019}. 
 Lavrinenko and colleagues~\cite{LavrinenkoPRB2019} demonstrate that polarization response of a y-aligned silver nanorod to a y-polarized probe pulse can be controlled by the presence of an x-polarized auxiliary pulse, see Fig.~\ref{fig1}. Owing to the shape resonance (selective coupling~\cite{andrews2019comprehensive}) of nanorods; x,y-polarized pulse can excite ``only" the x,y-aligned nanorods, respectively. Beyond studying the enhancement using a basic analytical model, which has widely been demonstrated to work very well for plasmonic path interference effects~\cite{Pelton2010OptExp,lovera2013mechanisms,TasginFanoBook2018}, Lavrinenko and colleagues also demonstrated the phenomenon via numerical solutions of 3D Maxwell equations. This scheme not only enhances the refractive index more than 2-orders of magnitude, but also results a canceled absorption at the enhancement frequency ---just as the same in its EIT counterpart~\cite{fleischhauer2005electromagnetically}. This phenomenon allows us to adjust the ``loss-free" refractive index of metamaterials without sacrificing  the probe signal to metallic losses.

In this paper, we show that plasmonic index enhancement is the dream-spouse for a photonic crystal. While other methods~\cite{reed2004silicon,almeida2004all,ctistis2011ultimate,yuce2013all} can tune the band gap size only about 2-3\%, the presented scheme can \textit{create} a large (e.g. 0.3 and 0.5 eV in Fig.~\ref{fig3}) from a \textit{nonexisting} gap only with an extremely small filling ratio~\cite{PSfilling}. One can achieve ``any particular value" of the band gap width between 0-0.3~eV for a y-polarized probe pulse~($E_1$) by merely arranging the {\it intensity}~($\sim |E_2|^2$) of the x-polarized auxiliary (pump) pulse between $E_2$=0-10$E_1$, working in the low intensity regime. $|E_1|^2$ is the intensity of the weak probe field. Moreover, response time is limited with the plasmon lifetime~($\sim$ fs~\cite{zoric2011gold}) and relaxation time as the scheme utilizes metal nanoparticles. In the case of silver nanorods, Ref.~\cite{LavrinenkoPRB2019} originally studies, a 100 fs linear photonic switching can be achieved using 30~fs auxiliary pulse, see Fig.~\ref{fig5}, where silver's plasmon lifetime we use is already 40 fs long. It is apparent that using a smaller plasmon lifetime material, such as platinum~\cite{zoric2011gold,yildiz2020plasmon}, should result much faster switching times~\cite{PSlargerdamping}.

Owing to the supreme index enhancement, PBG effects show up, see Figs.~\ref{fig3} and \ref{fig4}, via 6 dimer layers. This helps the miniaturization of PCs. The sample setup, Fig.~\ref{fig1}, presents a 750 nm long enhanced photonic crystal on a PC-waveguide chip employing CMOS-compatible noble metal (silver) materials.

We also present a demonstration of the phenomenon via numerical solutions of Maxwell equations (Fig.~\ref{fig4}). We kindly remark that our aim, in this {\it prominent} work, is merely presenting the first study for the utilization of plasmon index enhancement for PBG structures operating at room temperature.

\begin{figure}
\centering
\includegraphics[width=0.5\textwidth]{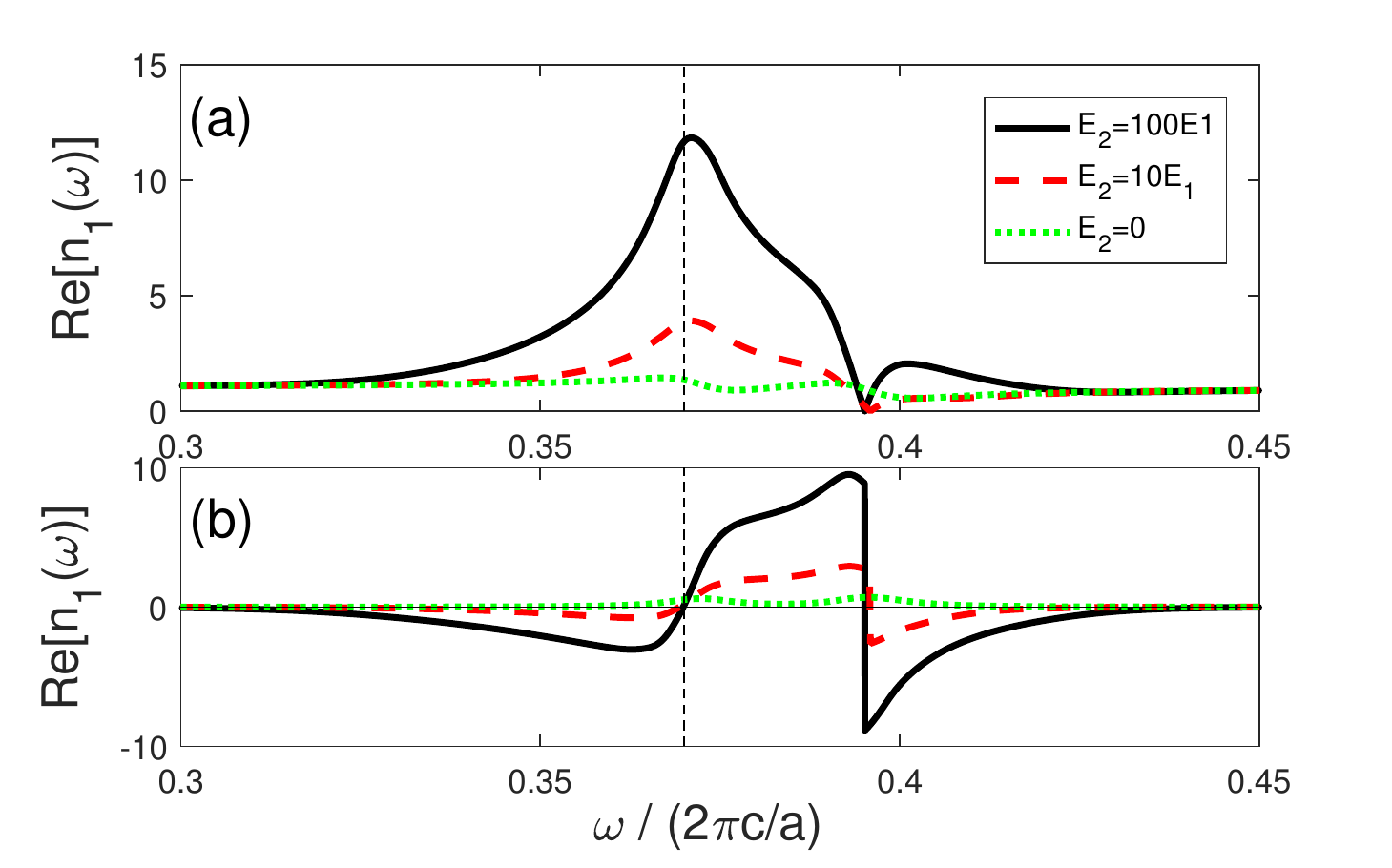} 
\caption{(a) Real and (b) imaginary parts of the refractive index. Index of the y-polarized probe pulse is controlled by the intensity of the auxiliary pulse~($E_2$). Index is both enhanced and real at $\omega_{\rm enh}$=0.37$\times2\pi c/a$. ($a$=200 nm) Without the periodicity, probe pulse is amplified, e.g., to the left of $\omega_{\rm enh}$. }
\label{fig2}
\end{figure}

\begin{figure}
\centering
\includegraphics[width=0.5\textwidth]{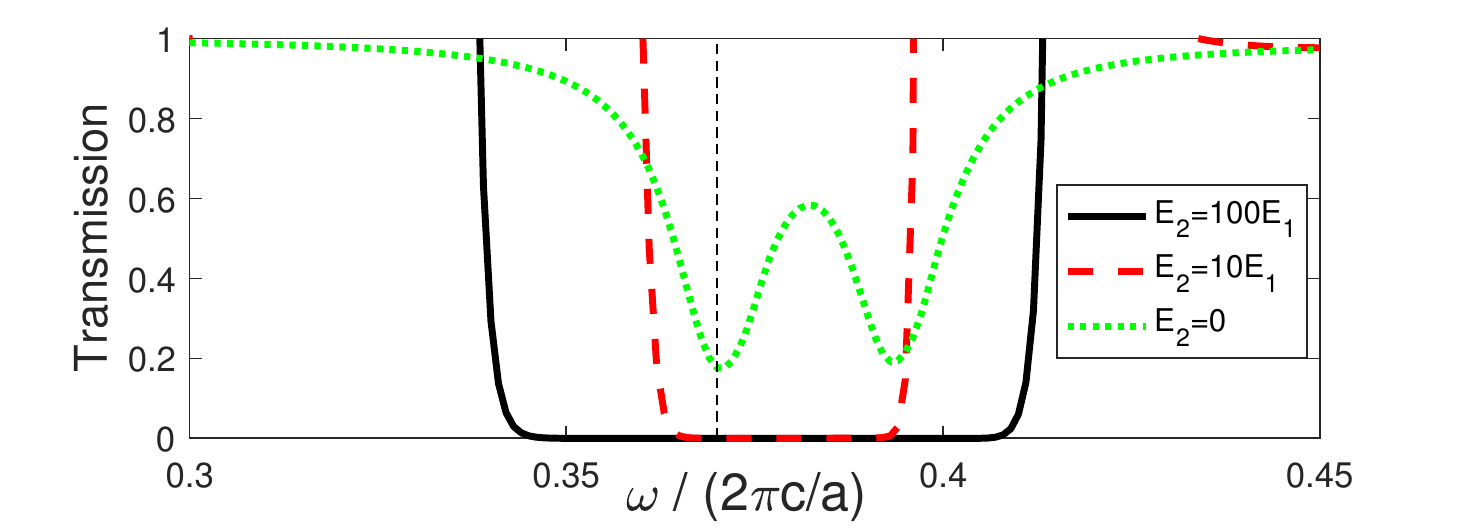} 
\caption{Probe transmission from only 6 layers of silver nanorods~(Fig.~\ref{fig1}). $E_2=100E_1$ and $10E_1$ display photonic band gaps even in the left of $\omega_{\rm enh}$=0.37 (vertical line) where there is a strong gain (Fig.~\ref{fig2}b). The width and existence of the band gap can be continuously-tuned by the auxiliary pulse intensity.  }
\label{fig3}
\end{figure}

\begin{figure}
\centering
\includegraphics[width=0.52\textwidth]{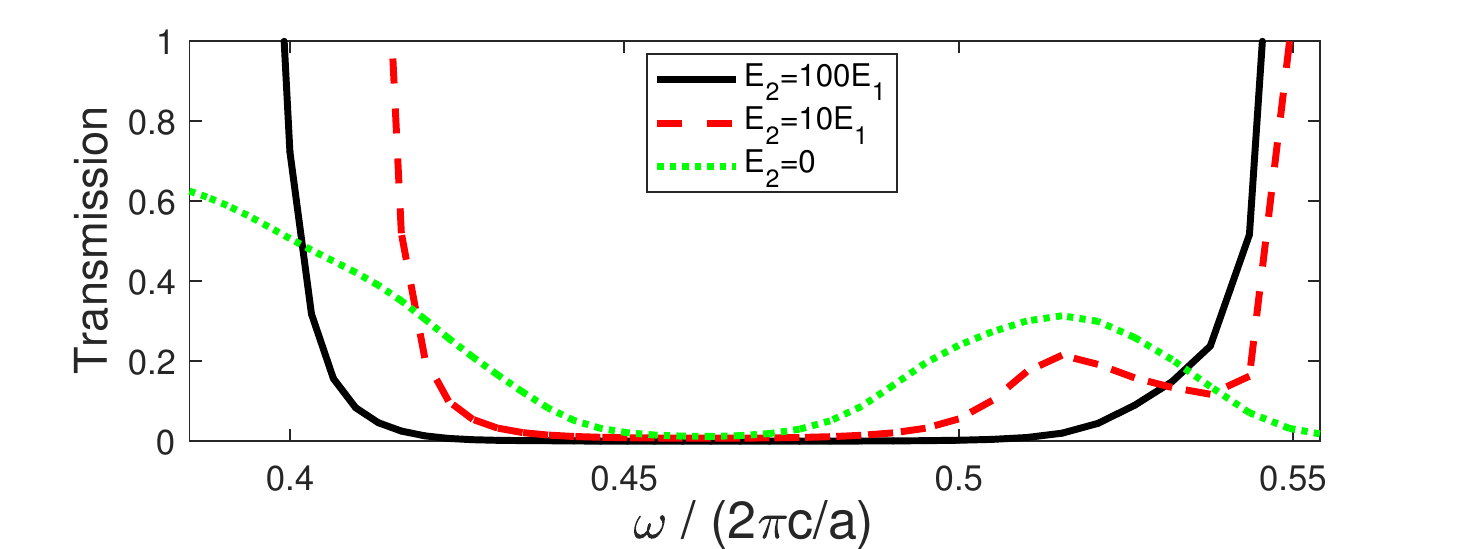} 
\caption{ FEM (COMSOL) simulations for a PC of 10-silver dimers. $a=200\:{\rm nm}$. Transmission demonstrates the PBG phenomenon clearly.  Location of the PBG naturally somewhat deviates from theoretical calculations as the index enhancement frequency is different.  For $E_2= 0$ transmission is low due to metallic losses (also no reflection). However, for $E_2=10,100E_1$ show strong reflection (due to PBG). 
}
\label{fig4}
\end{figure}

\begin{figure}
\centering
\includegraphics[width=0.5\textwidth]{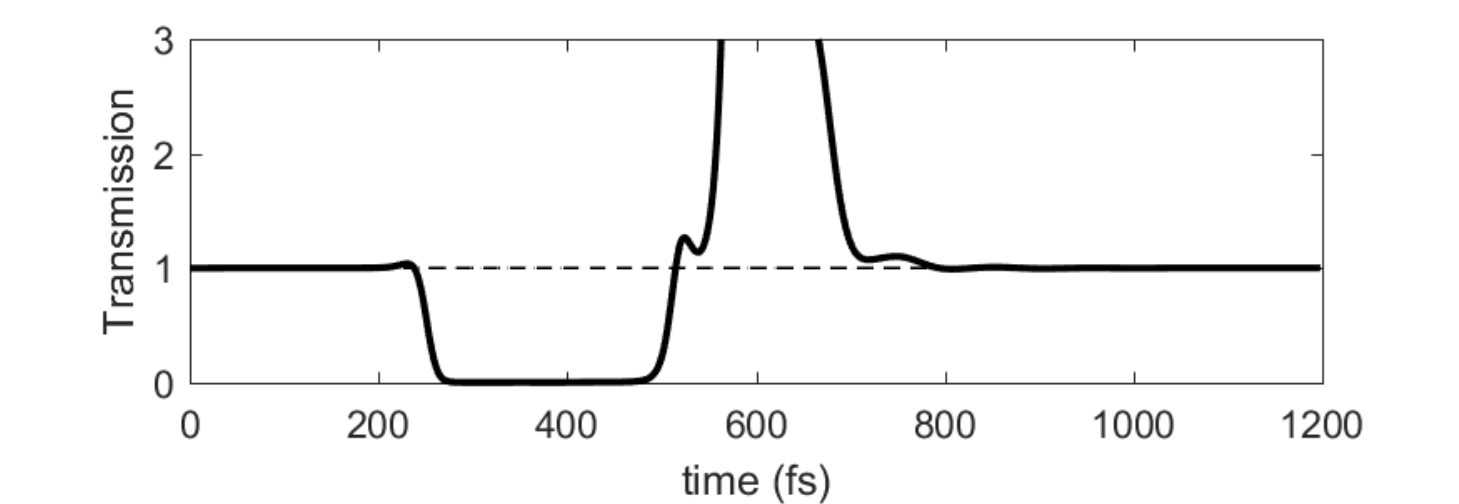} 
\caption{When the applied auxiliary field $E_2$ is an ultrashort pulse ($\Delta t$=40 fs), the crystal of silver dimers behaves as a linear photonic switch. The transmission (switch) along the x-direction is shutdown when the auxiliary pulse is on the y-aligned nanorods in Fig.~\ref{fig1}. The switch remains off for $\simeq 200$ fs. Length of this interval, limited  by the damping of the silver rod ($\tau \sim$ 38 fs), can be refined (shortened) substantially using a bad-quality nanostructure. For instance it is 1.3 fs in platinum~\cite{yildiz2020plasmon}. }
\label{fig5}
\end{figure}

{\small \bf Index-enhancement scheme.}--- Fig.~\ref{fig1} demonstrates a sample setup for implementing index-enhancement phenomenon for on-chip photonic operations. 6 layers of silver nanorod dimers (see the inset) switches the transmission of a $\rm y$-polarized probe pulse~($E_1$) of frequency $\omega$ via an x-polarized auxiliary (pump) pulse ($E_2$). When the auxiliary pulse is off, the refractive index of $20 \times 20 \times 60$ sized ($V_{\rm rod}=\pi\times4000 \: {\rm nm}^3$)  nanorods becomes exceedingly small for a density of 1 dimer in volume $V_{\rm site}$=200$\times$150$\times$200 ${\rm nm}^3$. Here 200 nm is the periodicity and 150$\times$200 corresponds to width and depth of the cavity in which silver dimers are placed~\cite{PSnotcrucial}. We note that these are rough calculations. Filling ratio of the index-enhanced layer, a single dimer in each one, is ~0.1\%. Thus, probe is transmitted with a relatively small absorption, i.e., 35\%. (Probe can be transmitted without any absorption as well, see below.) When the auxiliary pulse is turned on, with amplitudes $E_2=10E_1$ and $100E_1$, however, effective refractive index of the thin layers (10 nm length) is enhanced dramatically, see Fig.~\ref{fig2}a, thus turning ON a PBG of width $\sim$0.3 eV and $\sim$0.5 eV, respectively. See Fig~\ref{fig3}. The x-polarized auxiliary pulse couples only to the x-aligned silver nanorod. The x-aligned nanorod affects the response of the  y-aligned nanorod via a coupling through the hotspot appearing at the intersection of the two orthogonal rods. 
When the auxiliary pulse is present, y-polarized nanorod gives different polarization responses~($\chi_1=P_1/E_1$) for different values of the probe field $E_1$.

Dynamics of such a coupled system can be described by a basic model~\cite{LavrinenkoPRB2019,Cherenkov_Nanophotonics_2020}
\begin{eqnarray}
\ddot{x}_1+\gamma_1 \dot{x}_1 + \omega_1^2 x_1 - gx_2 = \tilde{E}_1(t), \label{x1}
\\
\ddot{x}_2+\gamma_2 \dot{x}_2 + \omega_2^2 x_2 - gx_1 = \tilde{E}_2(t), \label{x2}
\end{eqnarray}
which is very successful in demonstrating the Fano-like path-interference effects in plasmonic nanostructures~\cite{Pelton2010OptExp,lovera2013mechanisms,TasginFanoBook2018}. 
 Simulations in Ref.~\cite{LavrinenkoPRB2019} and the FEM simulations we carry out here, also demonstrate the appearance of the phenomenon. $\gamma_1=\gamma_2=0.026\omega_0$, $\omega_1=\omega_2=\omega_0$ and $g=0.06\omega_0^2$~\cite{LavrinenkoPRB2019} are the damping rates, resonances and the coupling between the two plasmon oscillations of the two nanorod, respectively. $\omega_0$ is the resonance of both nanorods. Here, only $\gamma_{1,2}$ is different from the one of Ref.~\cite{LavrinenkoPRB2019} since we determine the damping rates of nanorods from experimental dielectric function of silver. Actually, such a tidy choice of parameters is not necessary for this initial work which aims to present the implementation of the phenomenon for photonic technologies.  

For a monochromatic auxiliary pulse $E_2 \sim e^{-i\omega t}$, of the same frequency with the probe, the numerical solution for the susceptibility of the medium
\begin{equation}
\chi(\omega)=f\omega_0^2\frac{\delta_2+ge^{-i\phi}E_2/E_1}{\delta_1\delta_2-g^2}
\label{chi}
\end{equation} 
can be obtained~\cite{LavrinenkoPRB2019,Cherenkov_Nanophotonics_2020}. $\delta_i=\omega_i^2-\omega^2-i\gamma_i\omega$. The dimensionless oscillator strength $f$ is determined by setting nonenhanced susceptibility $\chi(\omega=\omega_0,g=0)=f/(\gamma_1/\omega_0)$~\cite{Cherenkov_Nanophotonics_2020} to $P_1/E_1$, with $P_1$ is the polarization density calculated for ellipsoid nanorods~\cite{LavrinenkoPRB2019,bohren2008absorption,Cherenkov_Nanophotonics_2020}, for 1 dimer in index-enhanced layer (20 nm), with volume $V_{\rm layer}$=20$\times$150$\times$200 ${\rm nm}^3$ in Fig.~\ref{fig1}. 

Since the auxiliary field, in our case an ultrashort pulse, has a broad frequency band, we take $\tilde{E}_2=\sum_\omega E_2 e^{-i\phi} \exp[{-(\omega-\Omega)^2/(\Delta\omega)^2}] e^{-i\omega t}$ pulse unlike Ref.~\cite{LavrinenkoPRB2019}. We solve Eqs.~(\ref{x1}-\ref{x2}) in the frequency domain for the Gaussian auxiliary pulse. $\Delta\omega=0.1\omega_0$. Fig.~\ref{fig2} plots the real and imaginary parts of the enhanced index~\cite{PSindexsign} for $E_2=100E_1$, $10E_1$ and $0$. We note that a double resonance for $E_2=0$ appears, because coupling between the two nanorods exists when the auxiliary pulse is turned off. 

In Fig.~\ref{fig2}b, we observe that imaginary part of the index becomes zero almost at the same frequency (vertical dashed-line) for $E_2=100E_1$ and $10E_1$, where the index becomes superiorly enhanced.

{\bf \small The phase plate.}--- Setup in Fig.~\ref{fig1} includes a critical apparatus; a phase plate. Eq.~(\ref{chi}) shows that the phase-difference between the auxiliary and probe pulse, $e^{-i\phi}$, plays an important role for the index enhancement to appear. The aux-probe phase-differences for the first and second dimer are not the same, since the probe does not have the same phase at different dimers (x-positions). Hence, when the first dimer is set to enhanced index (for y-polarized probe), the second dimer can possess a non-enhanced index; destroying the periodicity. Thus, we rephase the front of the auxiliary pulse with a phase plate either in the entrance of the aux port of the waveguide, or in front of (or inside) the auxiliary source. {\bf \small  Such a necessity for phase-tuning, most probably, is the reason large PBG changes cannot be observe in experiments with periodic nanoparticles~\cite{fan2015dynamic,he2019tunable,kang2019recent}}.

{\bf \small Photonic band gap.}--- In Fig.~\ref{fig3}, we present the transmission for a 6-layer index-enhanced PC of dimers with periodicity $a$=200 nm. It is striking that: while there exists a large amount of gain in the left of $\omega_{\rm enh}\simeq 0.37\times 2\pi c/a$ in Fig.~\ref{fig2}b, Fig.~\ref{fig3} shows that PGB, the 6-layer enhanced PC creates, can turn-OFF the transmission in a wide frequency region both for $E_2=100E_1$ and $10E_1$. We note that, in the figures we scale the frequencies by $\omega_a=2\pi c/a$ and momentum-vector by $k_a=2\pi /a$. In determining the natural band gap region, we first calculate the band diagram for the constant index $n=n(\omega_{\rm enh})$, then we choose periodicity $a$ accordingly as performed in Ref.~\cite{tasgin2007photonic}.

We also calculate the photonic band structure where a similar order $(e^{ik_I a})^6$  for the transmission appears. $k_I$ is the imaginary part of the wavevector-solution at $\omega_{\rm enh}\simeq$0.37.

{\bf \small Numerical Solutions of Maxwell’s Equations.}--- We also perform FEM (frequency domain) simulations using COMSOL as a proof of principle demonstration of the phenomenon. First, we obtain the y-component of the susceptibility $\chi$ by simulating the response of a single dimer which is coupled to the probe and auxiliary fields whose phases differ by $\pi$/2. We assume that the y-component of the electric field response of the dimer is directly proportional to the y-component of susceptibility.  We acquire the refractive index that has a similar form with the theoretical results depicted in Fig.~\ref{fig2}. Then, we input the calculated index values $n_y(x,y,z)$ back to 10 y-aligned nanoparticles separated by the lattice parameter $a$. Transmission properties of the system is depicted in Fig.~\ref{fig4}. The $a$ value is determined independently from the theoretical calculations~\cite{PSSSS}: we tested $a$ values from 150 nm to 300 nm. We chose $a$ such that the PBG matches the index enhancement frequency.

There exist one challenging situation FEM simulations made us realize. The probe field $E_1$ decays after each layer while propagating through the PC. This changes the $E_2/E_1$ ratio, hence, the periodicity of the refractive index. In such a setup while PBG effects are still clearly visible, high modulation depths may not be achieved. For circumventing this issue, we propose using another auxiliary wave. An x-polarized (additional) auxiliary wave illuminates the dimers along the z-axis~\cite{PSzdir} with $E_2 > |E_1^{(\rm aux)}| \gg |E_1|$. This way, $(E_1+E_1^{\rm (aux)})\hat{\bf y}$ creates an unchanged field along x-direction. This is also compatible with our simulations.

We were compelled to proceed with the 2-step approach, described above, in stead of simulating 10-dimers PC in a single step with the auxiliary fields. As the auxiliary pulse intensities would be much larger than the probe intensity, any remnant of aux fields reflected/diffracted  at the simulation boundaries do interfere with the probe field~\cite{comsolForum,PSFEMsimulation}.  This severely complicates the simulation. Our approach, advantageously, is not vulnerable against such complications.

{\bf \small PC operation regimes.}--- There can be several options for the operation regions in utilizing the system as a photonic switch. (a) PC can be operated at $\omega=\omega_{\rm enh}$, but transmission in the switch-on mode drops to 0.15 for this regime, see Fig.~\ref{fig3}a. (b) One can choose a better switch-on transmission 0.55 for $\omega\simeq 0.38$, since absorption in the switch-off mode is not important for this application. (c) Index-enhancement scheme is an active medium. It displays a substantial gain, e.g., between $\omega$=0.35-0.37, when the auxiliary (pump) field is set to $E_2=10E_1$. Thus, working, for instance, $\omega=0.36$ enhances the transmitted pulse several orders of magnitude for $E_2=10E_1$ via gain, but, suppresses the transmission when auxiliary is turned to $E_2=100E_1$ due to the band gap.  While we demonstrate the switch mechanism for quite different values of the auxiliary field, i.e., $E_2=100E_1$ and $10E_1$, an optimization can be carried out for smaller differences.

{\bf \small Response time.}--- In Fig.~\ref{fig5}, we present the time development of the transmitted pulse when a 40 fs, auxiliary pulse passes through the enhanced PC. We observe that, although the temporal width of the auxiliary pulse is 40 fs PBG remains off for about $\sim$200 fs. This limitation, actually, occurs due to two transient times. The decay time of the surface plasmon oscillations on silver, used in this work~\cite{LavrinenkoPRB2019,Cherenkov_Nanophotonics_2020}, is already about 40 fs. The switch time is also determined by the relaxation time, appearing due to the interaction between the two nanorods~\cite{PStransient}. After $t\simeq$600 fs, we observe the gain amplification of the transmission (it reaches $\sim 10^3$) during the PBG off regime. This transient also lasts $\sim$200 fs, but, increases the modulation depth between $t=$400 fs and 600 fs dramatically. When, for instance, platinum nanorods, of decay time $\sim$1 fs are used in the enhanced PC, much faster response times can be achieved. We remark that, a larger damping rate necessitates also a larger coupling $g$ for enhancement scheme to appear.

{\bf \small Periodicity on 2D and 3D.}--- Index enhancement scheme depends critically on the phase difference between the probe and the auxiliary pulses. When there is periodicity of nanodimers along the z-axis in Fig.~\ref{fig1}, there would appear no problem. Because, the phase of the auxiliary pulse is constant in the x-z plane. On working the periodicity along the propagation direction of the auxiliary pulse, y, however, one needs to take care of the different positions (so phases) of adjacent dimer arrays along the y-direction. Traveling along the y-directions, the first and the second dimer arrays show different phase-differences for the auxiliary and probe pulses. Index enhancement scheme can be different, even be demolished, in this case. This problem, however, can be circumvented in two ways. (i) One can choose the y-periodicity of the PC such that at a desired operation wavelength altering phases, $e^{i\phi_1}$ and  $e^{i\phi_2}$, overlap. (ii) A second approach: one could induce an appropriate phase dilation also in the probe field along the y-axis either in the waveguide input or just before PC input. One important point to stress is that in the enhancement scheme, index can be superiorly strengthened such that a single dimer array becomes sufficient for switching operations.

{\bf \small In summary,} we show that recently-explored index enhancement for plasmonic nanoparticles~\cite{LavrinenkoPRB2019} can be utilized in photonic crystal technologies for achieving 100\% PBG tuning. This is because, available adjustment of refractive index, using this scheme~\cite{LavrinenkoPRB2019}, is incomparably larger than the currently applied methods~\cite{reed2004silicon,almeida2004all,ctistis2011ultimate,yuce2013all}. Index enhancement, achieved with a ``vanishing loss", is such a large value (as in its EIT-counterpart~\cite{tasgin2007photonic,mustecapliouglu2005photonic,fleischhauer1992resonantly}) that using very small filling ratio for nanoparticles becomes sufficient. We also provide a proof of principle demonstration of the phenomenon via numerical solutions of Maxwell equations. 

Switching time of the PBG is determined by the decay rate of the metal nanoparticles, so, it is in the femtosecond regime. The gain nature enhances the modulation of the transmission. Vanishing loss at the enhancement frequency can be utilized as undamped switchable cavities at specific frequencies important for quantum optics (information) applications~\cite{reiserer2015cavity,scully2003extracting,hardal2015superradiant,zou2017quantum}.

Therefore, implementation of the plasmon-analog of refractive index enhancement (control) to photonic crystals provides a revolutionary instrument for photonic applications.

\begin{acknowledgments}
The FEM simulations are conducted solely by AKD, sophomore at Bilkent University. MET acknowledges support from TUBA GEBIP 2017 and TUBITAK 1001 No:117F118. AB and MET acknowledge support from TUBITAK 1001 No: 119F101.
\end{acknowledgments}

\bibliography{bibliography}

\begin{thebibliography}{50}
\bibitem [{PSf()}]{PSfilling}%
  \BibitemOpen
  \href@noop {} {}\bibinfo {note} {Filling ratio of a silver dimer in
  Fig.~\ref{fig1} is 0.1\%.}\BibitemShut {Stop}%
\bibitem [{PSl()}]{PSlargerdamping}%
  \BibitemOpen
  \href@noop {} {}\bibinfo {note} {Using a material of larger damping also
  necessitates a larger coupling strength $g$ between the two nanorods, thus a
  smaller relaxation time, for the enhancement scheme to be
  observable.}\BibitemShut {Stop}%
\bibitem [{PSn()}]{PSnotcrucial}%
  \BibitemOpen
  \href@noop {} {}\bibinfo {note} {Obviously, these are not crucial
  values!}\BibitemShut {Stop}%
\bibitem [{PSS()}]{PSSSS}%
  \BibitemOpen
  \href@noop {} {}\bibinfo {note} {We note that the TMM and FEM simulations are
  not to be compared.}\BibitemShut {Stop}%
\bibitem [{PSF()}]{PSFEMsimulation}%
  \BibitemOpen
  \href@noop {} {}\bibinfo {note} {For a realistic calculation of the
  transmission rate, simulation boundaries are chosen close to the dimers.
  Response of the dimers can reach as high as $10^5\times E_1$ in the
  simulations. Thus, the boundary reflections become at least comparable with
  the probe field, if not exceeding it.}\BibitemShut {Stop}%
\bibitem [{PSt()}]{PStransient}%
  \BibitemOpen
  \href@noop {} {}\bibinfo {note} {In a coupled system, transient time is
  determined by the minimum of the damping rates.}\BibitemShut {Stop}%
\end{thebibliography}

\end{document}